\documentclass[twocolumn,aps,pra,showpacs,showkeys,superscriptaddress,floatfix]{revtex4-1}
\usepackage[T1]{fontenc}
\usepackage[utf8]{inputenc}
\usepackage{amsmath}
\usepackage{amssymb}
\usepackage{graphicx}
\usepackage{esint}
\begin{document}
\title{Quantum sensing of rotation velocity based on Bose-Hubbard model}
\author{Che Jiang}
\author{Yaojie Zeng}
\author{Qi Qin}
\author{Zhirui Gong}
\email{gongzr@szu.edu.cn}

\author{Hongchen Fu}
\email{hcfu@szu.edu.cn}

\affiliation{College of Physics and Energy, Shenzhen University, Shenzhen 518060,
P. R. China}
\begin{abstract}
This work theoretically study the Bose-Hubbard model in a ring geometry
in a rotating frame. We obtain an effective Hamiltonian by using unitary
transformation, where the effect of the rotating reference frame is
introducing additional phases to the hopping constant. Within the
mean-field theory, the phase transition edge of the Bose-Hubbard model
not only depends on the particle numbers and the ring radius, but
also depends on the rotation velocity. Therefore, we propose a sensing
method of the rotation velocity using the phase transition edge of
the Bose-Hubbard model. At the exact phase transition edge where this
sensing method is most sensitive, the resolution depends on the rotation
velocity, the particle numbers and the ring radius, while is independent
of the parameters in the Bose-Hubbard model such as the hopping constant
and the on-site interaction. 
\end{abstract}
\pacs{03.65.Vf, 42.50.Ct, 03.67.-a}
\date{\today}

\maketitle

\section{Introduction}

Gyroscope can measure rotation velocity to get orientation information
and plays a crucial role in inertial navigation~\cite{King98} and
geophysics~\cite{Schreiber04}. Furthermore, a gyroscope with extreme
high precision can facilitate fundamental researches such as general
relativity~\cite{Jonsson07}, etc. Comparing to a classical mechanical
gyroscope using the principle of angular momentum conservation, the
gyroscope utilizing quantum effects has great advantages such as high
precision, high sensitivity, stable components and compact size~\cite{Armenise10}.
On the other hand, the optical and laser gyroscopes measure the rotation
velocity by sensing the shift of the interference fringes between
two split laser beams, which is known as the Sagnac effect~\cite{Post67,Arditty81}.
Such sensing scheme is firstly proposed and observed by using optical
interferometer such as ring laser interferometers~\cite{Macek63,Aronowitz71,Stedman97,Rowe99}
and optical fiber interferometers~\cite{Ezekiel82,Lefevre93,Dakin97}.
The similar interference concept is then generalized to atom interferometer
gyroscope (AIG) ,the first kind of quantum gyroscope, measuring the
change of the two-photon Raman transitions for the manipulation of
atomic wave packets~\cite{Barrett14}. Various AIGs are theoretically
proposed and experimentally implemented using particles such as neutrons~\cite{Werner79},
the electrons~\cite{Hasselbach93} and neutral atoms~\cite{Riehle91,Lenef97,Gustavson97,Gustavson00,Oberthaler96}.
The second kind of quantum gyroscope is nuclear magnetic resonance
gyroscope(NMRG), which measures the rotation velocity by detecting
the precession frequency of the nuclear magnetic moment in the rotating
frame~\cite{Noor17}. Recently, the third type of the quantum gyroscope
has been experimentally implemented in the nitrogen-vacancy in diamond,
where the Berry phase shifts in the NV electronic ground-state coherence
facilitate the sensing of the rotation velocity~\cite{Noor17}.

In order to improve the rotation velocity measurement, novel quantum
effects are introduced into the sensing scheme. One example is that
characteristics of the quantum system can be very sensitive at the
critical point of the quantum phase transition, resulting in the transition
edge sensor(TES)~\cite{Irwin05,Sachdev,Zhang08}. Although the TES
is originally proposed to detect single photon in a superconducting
system, it can also facilitate other sensing schemes. Yu-Han Ma et.
al. explored the dynamics of transvers field Ising model (TFIM) and
found the Loschmidt echo of TFIM is sensitive to the rotation velocity
at the critical point of quantum phase transition (QPT)~\cite{Ma17}.
The reason to achieve higher resolution is that the Loschmidt echo
possesses much more rapid decay around the critical point of QPT.
Determined by various coefficients of the QPT system, if the phase
boundaries between different phases varies when the whole system is
placed in a rotating frame, the sudden changes of the order parameters
guarantee not only the high resolution of the sensing of the rotation
velocity, but also the various sensing schemes as well.

Therefore, this work explores more sensing schemes of the rotation
velocity by measuring the QPT as long as it is directly affected by
the rotation velocity in a rotating frame. Comparing to the transvers
field Ising model where the rotating frame equivalently plays the
role of an external magnetic field, this work considers the Bose-Huddbard(BH)
model in a rotating reference frame due to its Bosonic nature and
following reasons. First, various phase boundaries exist between the
Mott insulator phases and the superfluid phases because of different
occupation numbers. When the unitary transformation is applied to
transfer the non-inertial reference frame to an inertial one, the
rotation introduces additional phases to the hopping constant between
the nearest neighbor sites. It eventually changes the order parameter
of BH model and modifies the phase boundaries. Therefore, the rotation
velocity is obtained by measuring the changes of the order parameter.
The second reason is that the order parameter changes dramatically
at the phase transition edges. A sensing scheme of rotation velocity
is proposed using the QPT of BH model. It is found that the resolution
reaches its maximum value at the phase transition edges. Additionally,
it only depends on the rotation velocity, the particle numbers and
the ring radius, while it is independent of the Bose-Hubbard model
such as the hopping constant and the on-site interaction. Thus this
work may shed light on the quantum gyroscope using the transition
edge sensors.

The paper is organized as follows. In section II, we introduce the
BH model of the ring system in rotating frame. The sensing scheme
of the rotation velocity is proposed in the Sec. III. In Sec IV, we
calculate the resolution of the sensing scheme.We conclude in section
V.

\section{The Bose-Hubbard Model in Rotating Reference Frame}

The Bose-Hubbard model is used to describe the phase transition between
the Mott insulator phase and superfluid phase at zero temperature,
which has been experimentally implemented in a cold atom system. The
Hamiltonian of the BH model in an inertial reference is introduced
as

\begin{equation}
\hat{H}=-t\sum_{\left\langle i,j\right\rangle }(\hat{a_{i}}^{\dagger}\hat{a_{j}}+h.c.)-\mu\sum_{i}\hat{n_{i}}+\frac{U}{2}\sum_{i}\hat{n_{i}}(\hat{n_{i}}-1)\label{eq:2-1}
\end{equation}
Where the $\hat{a_{i}}$ and $\hat{a_{i}}^{\dagger}$ are respectively
the annihilation and creation operators of bosons at $i-th$ site,
$t$ is the hopping constant between the nearest neighbor sites $\left\langle i,j\right\rangle $,
$U$ is the on-site interaction, $\mu$ is the chemical potential
and $\hat{n_{i}}=\hat{a_{i}}^{\dagger}\hat{a_{i}}$ is the particle
number operator. The system prefers Mott insulator phase for small
$t/U$ and prefers the superfluid phase for large $t/U$.

\begin{figure}[ptb]
\includegraphics[width=3.4in]{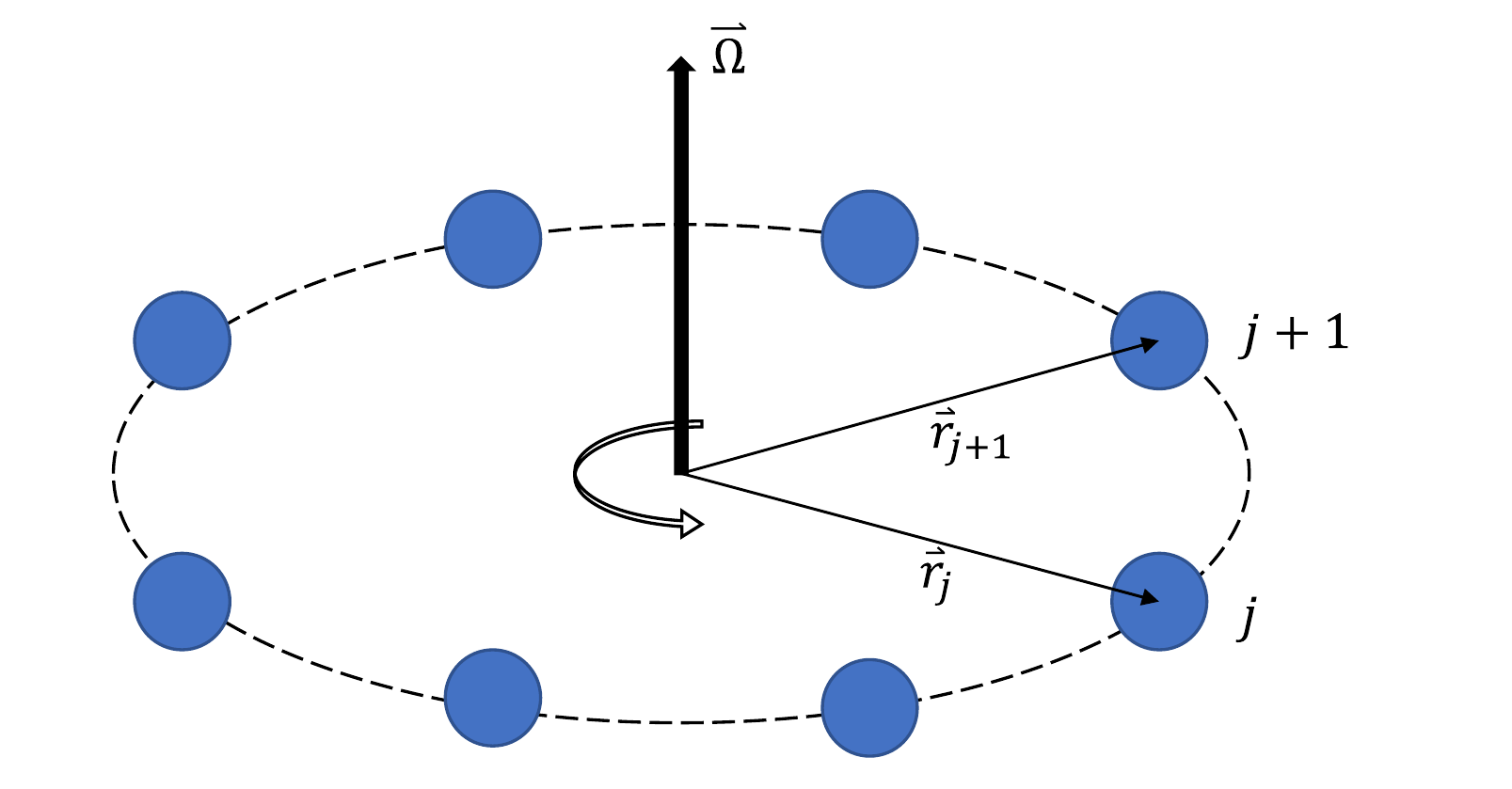}

\caption{(Color online). Schematic of the Bose Hubbard model in the rotating
reference frame with rotation velocity $\Omega$.}

\label{fig:fig1} 
\end{figure}

Now we consider the BH model in a rotating reference frame with rotating
angular velocity $\Omega$. Without loss of generality, the rotating
axis is along the z-axis as $\overrightarrow{\Omega}=\Omega\left(0,0,1\right)$.
In the inertial reference frame, the effect of the rotating reference
frame can be described by the following unitary transformation as

\begin{equation}
\hat{H}_{lab}=U^{\dagger}\left(t\right)\hat{H}U\left(t\right)+\left(i\hbar\frac{d}{dt}U^{\dagger}\left(t\right)\right)U\left(t\right),\label{eq:2-2}
\end{equation}
with the time-dependent unitary transformation matrix $U\left(t\right)=\exp\left(-\frac{it}{\hbar}\overrightarrow{\Omega}\cdot\overrightarrow{L}\right)$
with angular momentum $\overrightarrow{L}=\overrightarrow{r}\times\overrightarrow{p}.$
After the second quantization, the Hamiltonian in the laboratory reference
frame can be written as 
\begin{eqnarray}
\hat{H}_{lab} & = & -t\sum_{j}\left[\hat{a}_{j}^{\dagger}\hat{a}_{j+1}\exp\left(i\theta_{j}\right)+h.c.\right]\nonumber \\
 &  & -\mu\sum_{j}\hat{n}_{j}+U\sum_{j}\hat{n}_{j}\left(\hat{n}_{j}-1\right),\label{eq:2-3}
\end{eqnarray}
where$\theta_{j}=\intop_{r_{j+1}}^{r_{j}}d\vec{r}\cdot\overrightarrow{A}$
is the Peierls phases at the $j-th$ site and $\overrightarrow{A}=\frac{m}{\hbar}\overrightarrow{\Omega}\times\overrightarrow{r}$
is the velocity vector with particle mass $m.$ As we can see, the
effect of the rotating reference frame is introducing the site-dependent
Peierls phases at each site, which eventually will affect the QPT
of the BH model. In the sake of simplicity, we consider a ring geometry
of BH model with ring radius $r$ and rotates around the axis $\left(0,0,1\right)$
with rotation velocity $\overrightarrow{\Omega}$ (Shown in Fig. ~\ref{fig:fig1}).
In this sense, the Peierls phases become site-independent ones as
\begin{equation}
\theta\equiv\intop_{\overrightarrow{r}_{j+1}}^{\overrightarrow{r}_{j}}d\vec{r}\cdot\overrightarrow{A}=\frac{m}{\hbar}\Omega R\intop_{r_{j+1}}^{r_{j}}dr=\frac{2\pi R^{2}}{N}\frac{m}{\hbar}\Omega.\label{eq:2-4}
\end{equation}

In order to demonstrate how the rotation velocity affects the quantum
phase transition, the mean field theory is adopted where the order
parameter is the average of the annihilation operator $\left\langle \hat{a}_{j}\right\rangle \text{=\ensuremath{\psi}}$
on the ground state. In this sense, the hopping terms can be written
as

\begin{equation}
\hat{a}_{j}^{\dagger}\hat{a}_{j\pm1}\rightarrow\hat{a}_{j}^{\dagger}\left\langle \hat{a}_{j\pm1}\right\rangle +\left\langle \hat{a}_{j}^{\dagger}\right\rangle \hat{a}_{j\pm1}-\left\langle \hat{a}_{j}^{\dagger}\right\rangle \left\langle \hat{a}_{j\pm1}\right\rangle .\label{eq:2-5}
\end{equation}
Eventually the Hamiltonian become the summation of one-body Hamiltonian
$\hat{H}_{lab}=\sum_{j}H_{j}$ with

\begin{eqnarray}
H_{j} & = & -\mu n_{j}+Un_{j}(n_{j}-1)\nonumber \\
 &  & -2t\cos\theta\left(\hat{a}_{j}^{\dagger}\psi+\psi^{*}\hat{a}_{j}-\left|\psi\right|^{2}\right).\label{eq:2-6}
\end{eqnarray}

According to the Landau phase transition theory, the ground state
energy expanded to the forth-order of the order parameter $E=a_{0}+a_{2}\left|\psi\right|^{2}+a_{4}\left|\psi\right|^{4}$
have spontaneous symmetry breaking at the critical point $a_{2}=0.$
To analytically determine the ground state and the order parameter
simultaneously, we apply the forth-order perturbation theory and obtain
the following results

\begin{eqnarray}
a_{2} & = & 4t^{2}\cos^{2}\theta\left(\frac{n+1}{E_{n,n+1}}+\frac{n}{E_{n,n-1}}\right)\label{eq:2-7}
\end{eqnarray}

\begin{eqnarray}
a_{4} & = & 16t^{4}\cos^{4}\theta\left[\frac{\left(n+1\right)\left(n+2\right)}{E_{n,n+1}^{2}E_{n,n+2}}\right.\nonumber \\
 &  & +\frac{n\left(n-1\right)}{E_{n,n-1}^{2}E_{n,n-2}}-\frac{\left(n+1\right)^{2}}{E_{n,n+1}^{3}}-\frac{n^{2}}{E_{n,n-1}^{3}}\nonumber \\
 &  & \left.-\frac{n\left(n+1\right)}{E_{n,n+1}E_{n,n-1}^{2}}-\frac{n\left(n+1\right)}{E_{n,n+1}^{2}E_{n,n-1}}\right],\label{eq:2-8}
\end{eqnarray}
where $E_{n,m}=E_{n}^{(0)}-E_{m}^{(0)}$ is the energy difference
between the eigen-states with eigen-energies $E_{n}^{(0)}=-\mu n+Un(n-1)$
in the local limit ($t=0$).

The critical point $a_{2}=0$ determines the phase boundary as

\begin{equation}
\frac{1}{\widetilde{D}}=-\frac{n+1}{\widetilde{\mu}-2n}+\frac{n}{\widetilde{\mu}-2\left(n-1\right)}
\end{equation}
where $\widetilde{\mu}=\frac{\mu}{U}$, $\widetilde{D}=\widetilde{t}\cos\theta$,
$\widetilde{t}=\frac{t}{U}$. Above the critical point $\left(a_{2}>0\right)$
the system stays in Mott insulator phase with zero order parameter
$\psi=0$. When getting below the critical point $\left(a_{2}<0\right)$
the system stays in the superfluid phase with nonzero order parameter
$\psi=\sqrt{-a_{2}/\left(2a_{4}\right)}$, which can be experimentally
measured by the critical momentum in the transport measurement.

The rotation of the system affects the ground state of the system
and eventually shifts the critical point and the phase boundary. The
information of the rotation velocity is also contained in the change
of the order parameter, which is experimentally measurable. Therefore
by measuring the change of the order parameter, we can sense the rotation
velocity.

\section{Quantum sensing scheme of the rotation velocity}

\begin{figure}[ptb]
\includegraphics[width=3.4in]{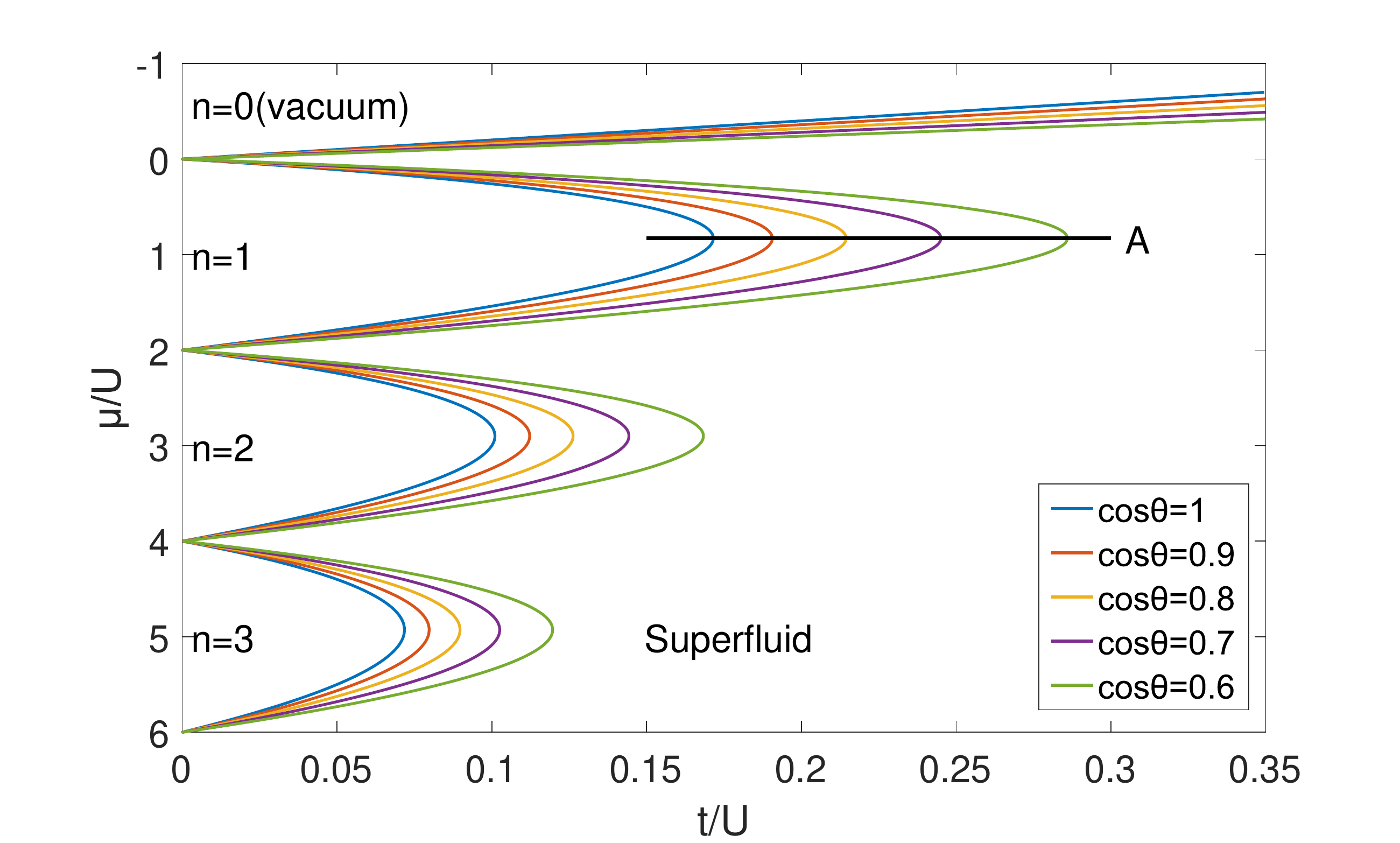}

\caption{(Color online). Phase diagram of Bose Hubbard model under mean field
method. Three lobes of different Mott insulators with $n=1,2,3$ together
with the vacuum and the superfluid phase are shown with respect to
different $\Omega$. The sensing loop $\widetilde{\mu}=\widetilde{\mu}\left(\widetilde{t},\Omega\right)$
is given by Eq.(9).}

\label{fig:fig2} 
\end{figure}

In this section, we propose a quantum sensing scheme of the rotation
velocity through the observation of the Bose-Hubbard Model's quantum
phase transition. The phase diagram of the Bose-Hubbard Model is analytically
obtained from Eq.(9), which is shown in Fig.~\ref{fig:fig2}. Here
the different phase boundaries correspond to different $\cos\theta$.
Since $\theta$ is proportional to the rotation velocity, $\cos\theta$
will decreases when the rotation velocity increases. As shown in Fig.~\ref{fig:fig2},
there are several lobes characterized by an integer value of $n$,
in which the ground state of system refers to the Mott insulator.
Outside the lobes the nonzero order parameter of the ground state
indicates the superfluidity of the system.


If the system in an inertial reference is originally prepared in a
superfluid state in the vicinity of phase boundary. Then the whole
system starts to rotate. The introduction of the rotation velocity
results in the expansion of the phase boundary as shown in Fig.~\ref{fig:fig2}.
The order parameter shall varies dramatically because of the phase
transition from superfluid to Mott insulator. The rotation velocity
can be derived by detecting the change of order parameter. Since the
rotation velocity only changes the hopping constant from $t$ to $t\cos\theta$,
the sensing loop from the superfluid to Mott insulator is a straight
line with the constant chemical potential $\mu$, which is also shown
as the solid line A in Fig.~\ref{fig:fig2}.


\begin{figure}[ptb]
\includegraphics[width=3.4in]{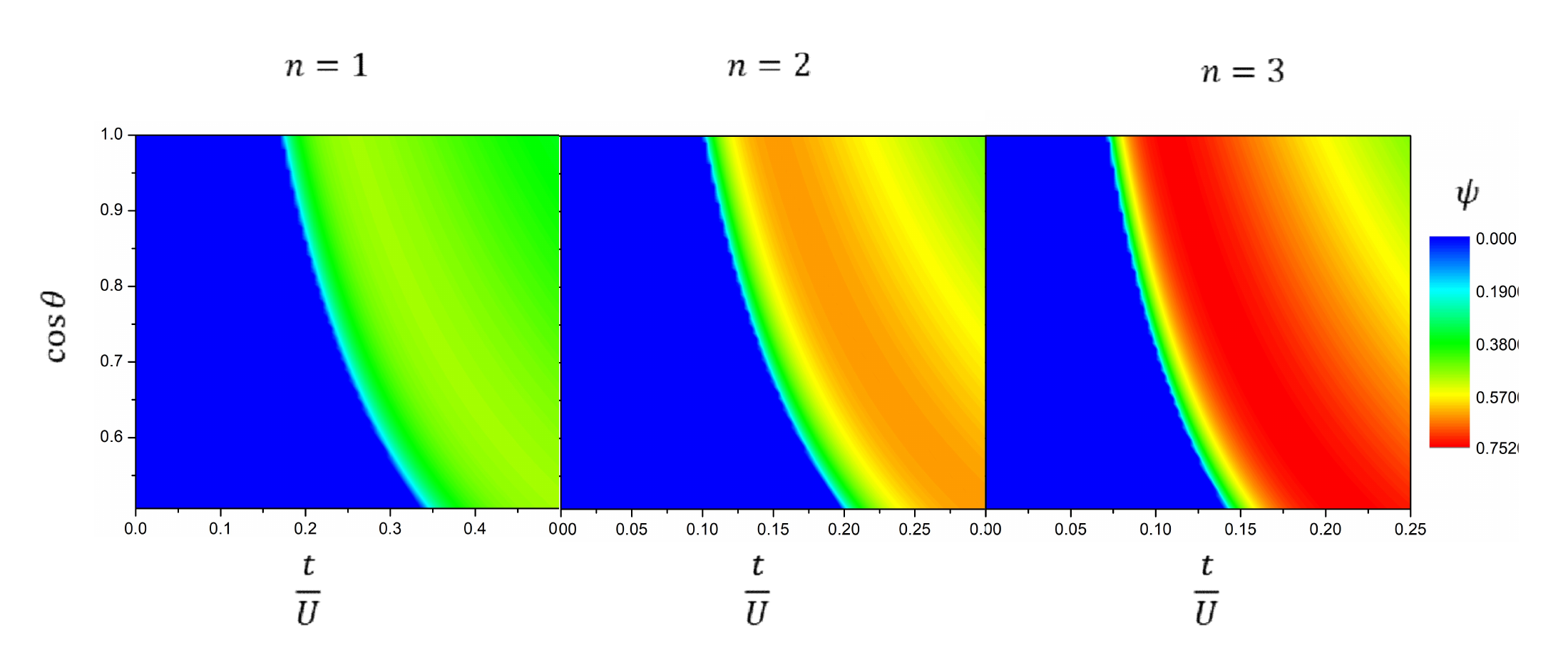}

\caption{(Color online) . Order parameters $\psi=\psi\left(\tilde{t},\Omega\right)$
versus $t/U$ and $\cos\theta$, with $\widetilde{\mu}=\widetilde{\mu}_{0}$,
$n=1$ (under different $\Omega$). }

\label{fig:fig3} 
\end{figure}

\begin{figure}[ptb]
\includegraphics[width=3.4in]{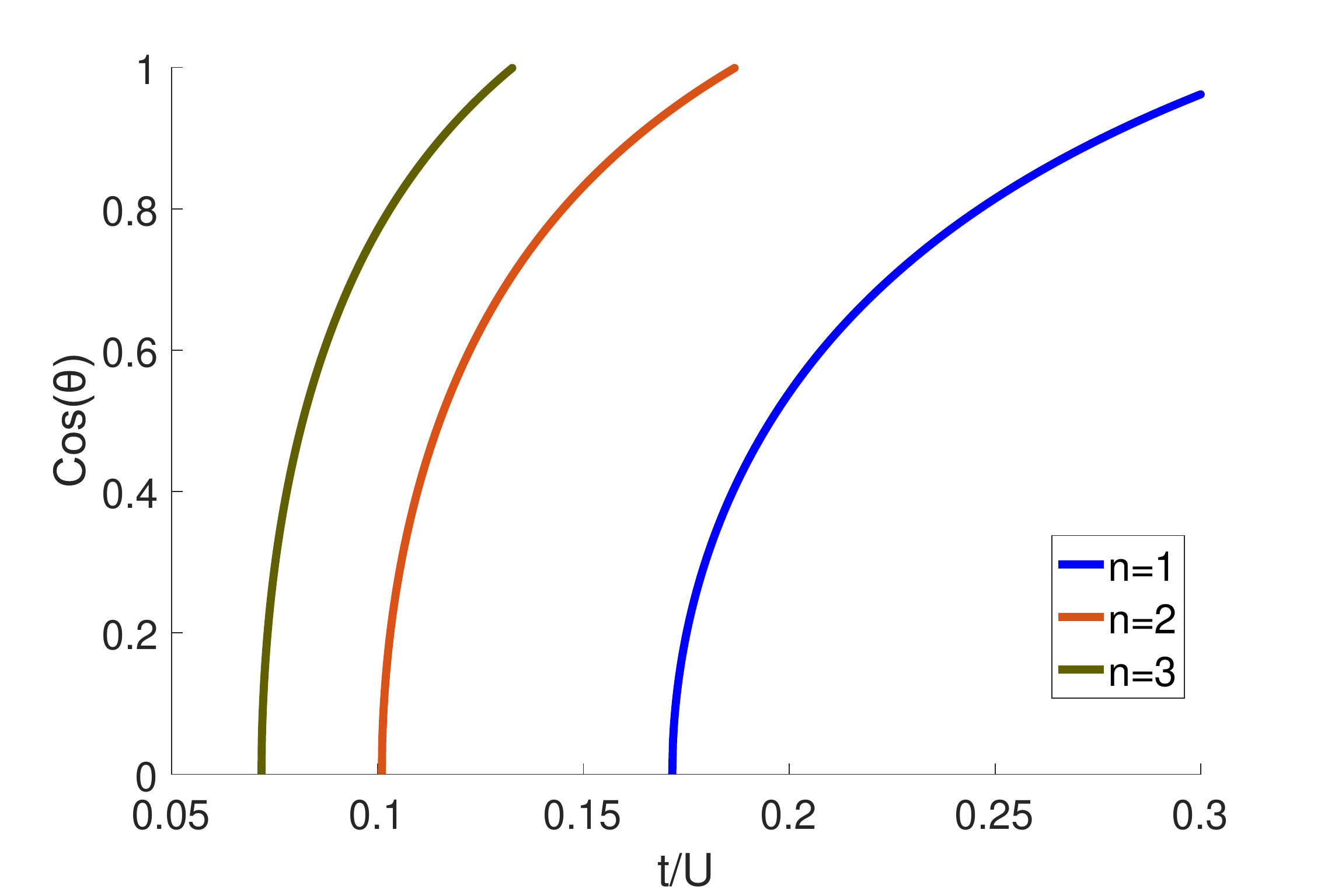}

\caption{(Color online) $\cos\theta$ varies with the change of $t/U$ for
different $n=1,2,3$ at $\mu=nU$, which is depicted as blue, red
and yellow solid lines.}

\label{fig:fig4} 
\end{figure}

The order parameters $\psi=\psi\left(\tilde{t},\Omega\right)$ along
line A versus the rotation velocity $\Omega$ and hopping constant
$t$ is shown in Fig.~\ref{fig:fig3}. The order parameter varies
dramatically in the vicinity of the phase boundary. By measurement
the variance of order parameter, the rotation velocity can be determined.
The variance of the order parameters is also determined by the $n$-th
slobe, the variance of the boundary is shown in Fig.~\ref{fig:fig4}.
Obviously, the variance of order parameter is more dramatic for higher
$n$, which indicates more efficient sensing scheme in the vicinity
of the phase boundary with higher $n$.


\section{The resolution of the quantum sensing scheme}

In order to characterize the efficiency of the quantum sensing scheme
based on the Bose Hubbard model, the resolution of the sensing scheme
with respect to the rotation velocity is obviously important. The
resolution can be defined as the full width at half(FWHM) of the change
in order parameters, which is defined as

\begin{equation}
\Delta=\psi\left(\widetilde{\mu},\widetilde{t},n,\Omega-\varDelta\Omega\right)-\psi\left(\widetilde{\mu},\widetilde{t},n,\Omega\right).
\end{equation}
The change in order parameters $\Delta$ is caused by the change in
rotation velocity($-\varDelta\Omega$, $0\leq\varDelta\Omega\leq\Omega$).
When the system is exactly prepared on the phase boundary ($a_{2}=0$),
the change in order parameters can be factorized as

\begin{equation}
\Delta=\kappa\left(\widetilde{\mu},n\right)\delta\left(\theta,\varDelta\theta\right),
\end{equation}
where 
\begin{eqnarray}
\kappa\left(\widetilde{\mu},n\right) & = & \frac{1}{\sqrt{2}}\left[\left(2n-\widetilde{\mu}-3\right)\right.\nonumber \\
 &  & \left.\left(2n-\widetilde{\mu}+1\right)\left(\widetilde{\mu}+2\right)^{3}\right]^{\frac{1}{2}}\nonumber \\
 &  & \left[-24n^{4}+48n^{3}\left(\widetilde{\mu}+1\right)\right.\nonumber \\
 &  & +2n\left(\widetilde{\mu}-10\right)\left(\widetilde{\mu}+1\right)\left(\widetilde{\mu}+2\right)\nonumber \\
 &  & +\left(\widetilde{\mu}+2\right)^{3}\left(\widetilde{\mu}+3\right)\nonumber \\
 &  & \left.-2n^{2}\left(\widetilde{\mu}\left(13\widetilde{\mu}+16\right)-8\right)\right]^{-\frac{1}{2}}
\end{eqnarray}

\begin{eqnarray}
\delta\left(\theta,\Delta\theta\right) & = & \cos\theta\sec\left(\theta-\varDelta\theta\right)\nonumber \\
 &  & \sqrt{\left(1-\cos\theta\sec\left(\theta-\varDelta\theta\right)\right)}
\end{eqnarray}
with $\theta=\gamma\Omega$, $\Delta\theta=\gamma\varDelta\Omega$
and $\gamma=\frac{2\pi mR^{2}}{N\hbar}.$ It should be pointed out
that since $\kappa\left(\widetilde{\mu},n\right)$ is independent
of the rotation velocity only $\delta\left(\theta,\Delta\theta\right)$
plays essential role in the resolution. For a fixed $\theta$ and
fixed $\Omega,$ $\delta\left(\theta,\Delta\theta\right)$ almost
exponentially increases when $\Delta\theta$ increases as shown in
Fig.~\ref{fig:fig4}. To obtain the resolution analytically, the
following approximate fitting formula is adopted as

\begin{eqnarray}
\delta\left(\theta,\Delta\theta\right) & \approx & \left(\sqrt{a\left(\theta\right)\Delta\theta}\right)\exp\left(-\sqrt{a\left(\theta\right)\Delta\theta}\right)
\end{eqnarray}
with one dependent fitting parameter $a\left(\theta\right).$


\begin{figure}[ptb]
\includegraphics[width=3.4in]{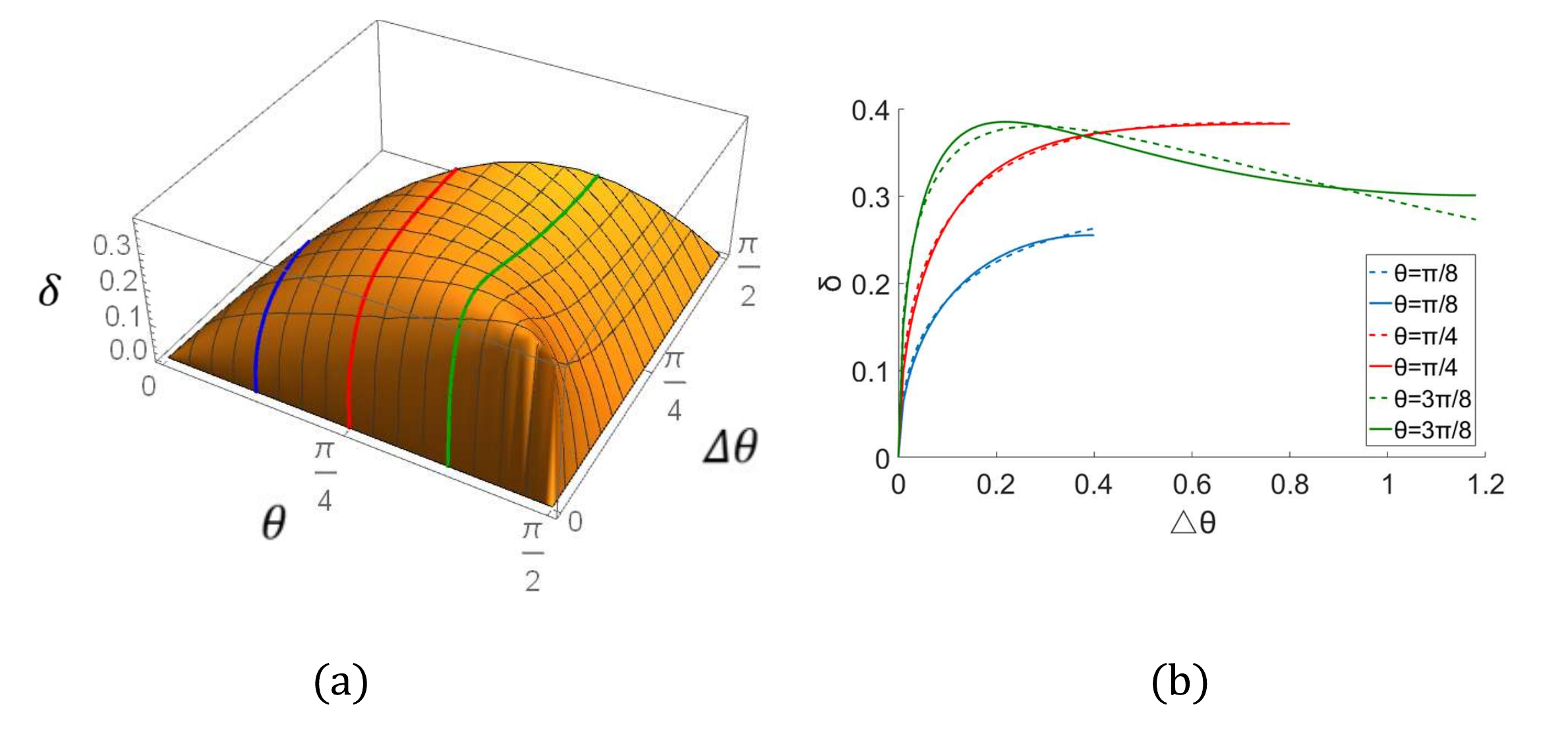}

\caption{(Color online) .Change of order parameter $\delta\left(\Omega,\Delta\Omega\right)$
versus the rotation velocity $\Omega$ and the change of rotation
velocity $\varDelta\Omega$. The solid lines in (b) denote the exact
change of order parameter and the corresponding dashed lines denote
the fitting one from Eq. (14). }

\label{fig:fig1-5} 
\end{figure}


In this sense, the $\delta\left(\theta,\Delta\theta\right)$ reaches
its maximum value $\delta_{m}=e^{-1}$ when $\Delta\theta=a\left(\theta\right)^{-1}$.
Here, not all maximum value can be exactly obtained because $\varDelta\theta\leq\theta$,
which gives rise to the critical point $\theta_{m}=\gamma\Omega_{m}\approx0.8603$
and the maximum value of $\delta\left(\theta,\Delta\theta\right)$
is

\begin{equation}
\delta_{m}=\begin{cases}
\left(\sqrt{a\left(\theta\right)\theta}\right)\exp\left(-\sqrt{a\left(\theta\right)\theta}\right) & \Omega<\Omega_{m}\\
e^{-1}, & \Omega\geq\Omega_{m}
\end{cases}
\end{equation}
Since the resolution defined as full width at half(FWHM) of the change
in order parameters, the resolution $\varepsilon$ is by the following
equation $\delta\left(\theta,\varepsilon\right)=\frac{\delta_{m}}{2}$
which is

\begin{equation}
\varepsilon=\begin{cases}
a\left(\theta\right)^{-2}ProductLog\left[-\frac{1}{2}\left(\sqrt{a\left(\theta\right)\theta}\right)\exp\left(-\sqrt{a\left(\theta\right)\theta}\right)\right]^{2}, & \Omega<\Omega_{m}\\
a\left(\theta\right)^{-2}ProductLog\left[-\frac{1}{2e}\right]^{2}, & \Omega\geq\Omega_{m}
\end{cases}
\end{equation}
Here the $ProductLog[z]$ is the solution of the equation $z=xe^{x}.$

The numerical calculation is shown in Fig.6. The red line and blue
line represents respectively the rigorous one by solving $\delta\left(\theta,\varepsilon\right)=\frac{\delta_{m}}{2}$
and the fitting one from Eq. 16. As shown in Fig. 6, the resolution
increases when the rotation velocity increases. The approximate resolution
fits the rigorous one very well.

\begin{figure}[ptb]
\includegraphics[width=3.4in]{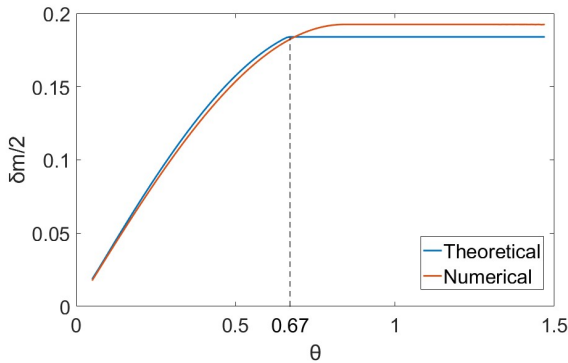}

\caption{(Color online) .Change of rotation velocity when change of order parameters
is FWHM,Theoretical function$\varepsilon=\frac{1}{4}\exp\left(-2\gamma\right)\cot\left(\Omega\right)$
fit well when $\theta\in\left[0.5,1.1\right]$}

\label{fig:fig1-6} 
\end{figure}


\section{CONCLUSION}

This work develops a Bose-Hubbard(BH) model in a rotating frame into
consideration due to its Bosonic nature. One reason to choose the
BH model is that various phase boundaries between the Mott insulator
phase and the superfluid phase emerges due to different occupation
numbers. When the unitary transformation is applied to transfer the
non-inertial reference frame to an inertial one, the rotation introduces
additional phases to the hopping constant between the nearest neighbor
sites. It eventually changes the order parameter of BH model, resulting
in the changed phase boundary. Therefore it is feasible to obtain
the rotation velocity by measuring the changes of the order parameter.
Another reason to choose the BH model is that the dramatic change
order parameter at the phase transition edges can improve the sensitivity.
We propose a sensing scheme of rotation velocity using the QPT of
BH model. We find that sensitivity reaches the maximum value at the
phase transition edges. Additionally, this sensitivity only depends
on the rotation velocity of the rotating reference, the particle numbers
and the ring radius, and it is independent of those parameters of
the Bose-Hubbard model such as the hopping constant and the on-site
interaction. This work may shed light on the quantum gyroscope using
the phase transition edges.

\begin{acknowledgements} The work is supported by National Natural
Science Foundation of China (Grant No. 12175150) and the Natural Science
Foundation of Guang-dong Province (Grant No. 2019A1515011400). \end{acknowledgements}


\end{document}